\documentclass[11pt]{article} 
\usepackage{epsfig}
\usepackage{natbib}
\begin{document}
%******************************************************************
%\baselineskip=.33in
%******************************************************************

%***********************
\newcommand{\be}{\begin{equation}}
\newcommand{\ee}{\end{equation}}
\newcommand{\bea}{\begin{eqnarray}}
\newcommand{\eea}{\end{eqnarray}}
\newcommand{\da}{\dagger}
\newcommand{\dg}[1]{\mbox{${#1}^{\dagger}$}}
\newcommand{\hlf}{\mbox{$1\over2$}}
\newcommand{\lfrac}[2]{\mbox{${#1}\over{#2}$}}
\newcommand{\scsz}[1]{\mbox{\scriptsize ${#1}$}}
\newcommand{\tsz}[1]{\mbox{\tiny ${#1}$}}
%***************

\begin{flushright} 
% gr-qc/0411095 
\end{flushright} 

\begin{center}

\Large{\bf Lunar Laser Ranging Science\rm\footnote{Presented at 14th International Workshop on Laser Ranging, San Fernando, Spain, June 7-11, 2004; conference webpage at {\tt http://www.roa.es/14workshop-laser/}}}

\vspace{0.4in}

\normalsize
\bigskip 

James G. Williams, Dale H. Boggs, Slava G. Turyshev, and J. Todd Ratcliff 

\normalsize
\vskip 15pt

{\it{Jet Propulsion Laboratory, California Institute of  Technology,\\
Pasadena, CA 91109, U.S.A.}} 
\footnote{Email: {\tt James.G.Williams@jpl.nasa.gov, Dale.H.Boggs@jpl.nasa.gov, 
Slava.G.Turyshev@jpl.nasa.gov, J.Todd.Ratcliff@jpl.nasa.gov }}
 
%***********************************************
% \today
%**************************************************

\vspace{0.25in}
%\bigskip 

\end{center}

%********************************************************************
%\baselineskip=.33in
%******************************************************************

\begin{abstract}
Analysis of Lunar Laser Ranging (LLR) data provides science results: gravitational physics and ephemeris information from the orbit, lunar science from rotation and solid-body tides, and Earth science.  

{\it Science from the orbit:}  Sensitive tests of gravitational physics include the Equivalence Principle, limits on the time variation of the gravitational constant $G$, and geodetic precession.  The equivalence principle test is used for an accurate determination of the parametrized post-Newtonian (PPN) parameter $\beta$.  Lunar ephemerides are a product of the LLR analysis used by current and future spacecraft missions.  The analysis is sensitive to astronomical parameters such as orbit, masses and obliquity.  The dissipation-caused semimajor axis rate is 37.9 mm/yr and the associated acceleration in orbital longitude is $-25.7 ''/$cent$^2$, dominated by tides on Earth with a 1\% lunar contribution. 
 
{\it Lunar science:}  Lunar rotational variation has sensitivity to interior structure, physical properties, and energy dissipation.  The second-degree lunar Love numbers are detected; $k_2$ has an accuracy of 11\%.  Lunar tidal dissipation is strong and its $Q$ has a weak dependence on tidal frequency.  A fluid core of about 20\% the Moon's radius is indicated by the dissipation data.  Evidence for the oblateness of the lunar fluid-core/solid-mantle boundary is getting stronger.  This would be independent evidence for a fluid lunar core.  Moon-centered coordinates of four retroreflectors are determined.  

{\it Earth science:}  Station positions and motion, Earth rotation variations, nutation, and precession are determined from analyses.  

{\it Future:}  Extending the data span and improving range accuracy will yield improved and new scientific results.  Adding either new retroreflectors or precise active transponders on the Moon would improve the accuracy of the science results.  

\end{abstract}
\vspace{0.15in}

%\newpage

% main text
%*********************1) INTRODUCTION
\section*{Introduction}
A 13th Workshop paper gave a review of lunar science and gravitational physics results plus an extensive comparison of the dynamical and analysis experiences for the Moon and Earth \citep{Williams_Dickey_2003}.  This present paper updates JPL lunar science (Section~\ref{sec:lunar_s}) and gravitational physics (Section~\ref{sec:lunar_orbit}) results and discusses future Lunar Laser Ranging (LLR) (Section~\ref{sec:concl}).  For recent results from the Paris Observatory analysis center see paper by \citet{Chapront_etal_2003}.

\section{Lunar Science}
\label{sec:lunar_s}
  
LLR is important for understanding the properties of the Moon.  Early LLR data were important for determining gravitational harmonics and moment differences and discovering free librations and strong rotational energy dissipation.  Now the accurate data of recent years permits the Moon's interior properties to be investigated.   

\subsection{Tides on the Moon}
The elastic response of the Moon to tidal forces is characterized by Love numbers.  The second-degree tides are strongest and these tidal displacements are sensitive to the second-degree Love numbers $h_2$ and $l_2$ while the rotation is sensitive to the potential Love number $k_2$.  The amplitudes of the two largest monthly terms are both about 9 cm.  In practice, LLR solutions are more sensitive to $k_2$.  A solution, with data from 1970 to 2003, gives $k_2 = 0.0227\pm0.0025$ and $h_2 = 0.039\pm0.010$ \citep{Williams_Boggs_Ratcliff_2004}.  $l_2$ was fixed at a model value of 0.011.  For comparison, there is an orbiting spacecraft determination of the lunar Love number $k_2 = 0.026\pm0.003$ determined from tidal variation of the gravity field \citep{Konopliv_etal_2001}. 
 
Accurate Love numbers are valuable because they give information on the elastic properties of the lunar interior.  For comparison with the solution results, model calculations have been made for lunar Love numbers.  Love number calculations start with an interior model \citep{Kuskov_Kronrod_1998} which is compatible with seismic P- and S-wave speeds deduced from Apollo seismometry.  There is little seismic information below 1100 km and the seismic speeds have to be extrapolated into the deeper regions above the core.  The 350 km radius of the fluid iron core was adjusted to match the LLR-determined $k_2$ and small adjustments were made to the densities to satisfy mass and moment constraints.  In addition to matching the above $k_2 = 0.0227$, the model calculations give $h_2 = 0.0397$ and $l_2 = 0.0106$.  

The LLR solutions are also sensitive to tidal dissipation \citep{Williams_etal_2001}.  In general, the specific dissipation $Q$ depends on frequency.  In the above solutions, the whole-Moon monthly tidal $Q$ is found to be $33\pm4$.  For $k_2 = 0.0227$ the power-law expression for tidal $Q$ as a function of tidal period is determined to be $33({\rm Period}/27.212 {\rm d})^{0.05}$ so the $Q$ increases from 33 at a month to 38 at one year.  At tidal frequencies the Moon exhibits strong dissipation.  

\subsection{A Molten Lunar Core}

Evidence for a distinct lunar core comes from the moment of inertia 
\citep{Konopliv_etal_1998}, the induced dipole moment 
\citep{Hood_etal_1999}, and Lunar Laser Ranging.  LLR analyses indicate that the core is fluid and the detection of this molten core is a major accomplishment of the LLR effort.  This is a small dense core, presumably iron rich with elements such as sulfur which lower the melting point.  

In addition to strong tidal dissipation, the lunar rotation also displays a strong source of dissipation which is compatible with a fluid core \citep{Williams_etal_2001}.  This source of dissipation arises from the fluid motion with respect to the solid mantle at a fluid-core/solid-mantle boundary (CMB).  With the aid of \citet{Yoder_1995} turbulent boundary layer theory these dissipation results give a $1\sigma$  upper limit for radius of 352 km for a pure Fe core or 374 km for a fluid Fe-FeS eutectic \citep{Williams_etal_2001}.  Upper limits are used because any topography on the CMB or the presence of an inner core would tend to decrease the inferred radius.  More recent solutions find a somewhat stronger fluid core dissipation torque.  

The detection of the oblateness of the fluid-core/solid-mantle boundary would be independent evidence for the existence of a liquid core.  Fluid flow along an oblate boundary exerts torques on both the fluid and the overlying solid Moon.  In recent years rotation evidence for an oblate boundary has been strengthening \citep{Williams_Boggs_Ratcliff_2004}.  In the solutions core oblateness and $k_2$ anticorrelate resulting in a smaller $k_2$ value than previous spherical core solutions gave.  The core oblateness is expected to be the next major LLR lunar science result.  

The internal structure and material properties of the Moon must be deduced from external evidence and the deepest regions are the most elusive.  In order to determine the variety of permissible interior structures and properties, a large number of models have been generated which satisfy, within measurement uncertainties, four lunar quantities: the mean density, the moment of inertia's measure of mass concentration toward the center, the $k_2$ elastic response to solid-body tides, and tidal dissipation $Q$ \citep{Khan_etal_2004}.  Typically, the central regions of the acceptable models have a higher density core which can take several forms such as completely solid, completely fluid, and a solid inner core within a fluid outer core.  

Learning whether the Moon has a small solid core inside a liquid outer core is a future possibility.  There should be signatures in the rotation data if that is the case, but they could be small.  
  
\subsection{Positions on the Moon}

A return to the Moon with robotic and manned spacecraft is now projected.  This presents an opportunity for new retroreflectors and optical transponders to go to the Moon.  The LLR retroreflectors have the most accurately known positions on the Moon.  A small number of accurate positions on the Moon are serving as control points for lunar geodesy \citep{Davies_Colvin_Meyer_1987,Davies_etal_1994,Davies_Colvin_2000} and it is hoped that future missions enable this network of accurate control points to be expanded.

\section{Science from the Lunar Orbit}
\label{sec:lunar_orbit}

Einstein's general theory of relativity has proved remarkably successful.  Nonetheless, this is a time for improved tests of gravity.  In physics there is an expectation that a theory of gravity can be found which is compatible with the quantum theories of the stronger forces.  Among the most promising extensions of relativistic gravity beyond general relativity are the scalar-tensor theories.  These theories can give small violations of the equivalence principle as well as a time-varying gravitational ``constant,'' two quantities that LLR determines well. 
 
Different aspects of metric theories of gravity are described with parametrized post-Newtonian (PPN)  $\beta$ and $\gamma$  parameters.  These PPN parameters have a unit value for general relativity, but a deviation from unity at levels of $10^{-5}$ to $10^{-7}$ has been predicted by \citet{Damour_Nordtvedt_1993} and \citet{Damour_Piazza_Veneziano_2002}.  

The great stability of the lunar orbit allows LLR to use the orbital motion to make accurate tests of gravitational physics.  A discussion follows of LLR tests of the equivalence principle, the implication for PPN  $\beta$, and variation of the gravitational constant.  The following new LLR solution results used LLR data through April 2004 \citep{Williams_Turyshev_Boggs_2004}.  

\subsection{Equivalence Principle}

The Equivalence Principle is a foundation of Einstein's theory of gravity.  The LLR analysis tests the Equivalence Principle by examining whether the Moon and Earth accelerate alike in the Sun's field.   \citet{Nordtvedt_1968,Nordtvedt_1970} gave theoretical analyses of the effects of a violation of the Equivalence Principle.  For the Earth and Moon accelerated by the Sun, if the equivalence principle is violated the lunar orbit will be displaced along the Earth-Sun line, producing a range signature having a 29.53-day period.  

The LLR test of the Equivalence Principle shows that the Earth and Moon are accelerated alike by the Sun's gravity with $\Delta$acceleration/acceleration of $(-1.0\pm1.4)\times10^{-13}$.  This solution corresponds to a $(2.8\pm4.1)$ mm $\cos D$ signature in the lunar distance.  

A violation of the Equivalence Principle might depend on composition or the strength of the gravitational attraction within a finite body (gravitational self-energy).  The former was tested in the laboratory by Adelberger \cite{Adelberger_2001} with an uncertainty similar to the LLR result.  The latter requires large bodies such as the Moon or planets and it depends on PPN $\beta$  and $\gamma$.  Combining the LLR result, the laboratory composition result, and the recent Cassini time delay test of  \citep{Bertotti_Iess_Tortora_2003}, one derives  $\beta - 1 = (1.2\pm1.1)\times10^{-4}$.  This is the strongest limit on PPN $\beta$  to date and is not significantly different from the unit value of general relativity.  

\subsection{Does the Gravitational Constant Vary?}

Einstein's general theory of relativity does not predict a variable gravitational constant $G$, but some other theories of gravity do.  A changing $G$ would alter the scale and periods of the orbits of the Moon and planets.  LLR is sensitive to $\dot{G}/G$ at the 1 AU scale of the annual orbit about the Sun \citep{Williams_etal_1996}.  No variation of the gravitational constant is discernible, with $\dot{G}/G=(4\pm9)\times10^{-13}$~yr$^{-1}$.  This is the most accurate result published to date.  The uncertainty corresponds to 1.2\% of the inverse age of the Universe.  The scale of the solar system does not share the cosmological expansion.  The sensitivity of changing $G$ depends on the square of the LLR time span so significant improvements are expected when future data accumulate.  

\subsection{Geodetic Precession}

The geodetic precession of 19 mas/yr is a relativistic effect from the annual motion of the Earth-Moon system around the Sun.  From the precession of the lunar orbit, LLR has provided the only accurate determination of the geodetic precession to date.  The latest LLR result is $K_{gp} = -0.0019\pm0.0064$, where the quantity $K_{gp}$ gives the relative deviation of the geodetic precession from the general relativity value, so the uncertainty corresponds to 0.12 mas/yr.  Correlation is high with the lunar Love number $k_2$ and the core oblateness.  

Two objectives of the Gravity Probe B (GP-B) mission are to make accurate measurements of the Lense-Thirring effect and the geodetic precession using very precise gyroscopes.  For the Moon, the Lense-Thirring effect, a gravitomagnetic frame dragging by a spinning body, causes a very small precession rate which is too small for current measurement.  LLR is sensitive to the orbital counterpart of the gravitomagnetic effect.  If GP-B fulfills its objectives it will produce a more accurate geodetic precession test than LLR.  

\subsection{Other Relativistic Gravity Tests}

There is not a broadly agreed upon theory of relativistic gravity to replace Einstein's general theory of relativity.  Consequently, many alternatives have been proposed and there will be more in the future.  To be realistic any of these theories must be compatible with the tests of relativity that are provided by LLR and planetary ranging.  We can expect that LLR will continue to play an important role in winnowing out some of these theories.  

Dark matter and dark energy are among the more exotic aspects of modern astronomy.  The mysterious dark matter is known from its gravitational pull and it is more abundant than normal matter.  A search for a dark matter equivalence principle effect has been reported by \citet{Nordtvedt_Mueller_Soffel_1995} and a test for a preferred frame effect has been done by \citet{Mueller_Nordtvedt_Vokrouhlicky_1996}.  Dark energy, which causes the expansion of the universe to accelerate, is another surprising discovery.  

\subsection{Lunar Ephemeris}

Analysis of the LLR data is used to generate the lunar ephemeris which, in combination with the planetary ephemerides, is used to navigate interplanetary spacecraft.  The lunar and planetary ephemeris is available at {\tt http://ssd.jpl.nasa.gov/}.  The lunar rotation, the physical libration, is also part of the ephemeris files.  An accurate lunar ephemeris is critical for future missions to the Moon and the physical librations are critical to landing on the Moon, navigation on the Moon, and observations from the Moon. 

\subsection{Modeling}

To analyze LLR data with a lengthening span and high accuracy, the data analysis models and programs must be improved. \citet{Standish_Williams_2003} describe the lunar and planetary numerical integrator used to generate past ephemerides.  Many small modeling improvements would benefit analyses of current and future data; an overview is given by \citet{Williams_Turyshev_Murphy_2004}.  A new set of changes to the integration program has been started.  

\section{Current Interests and Future Possibilities}
\label{sec:concl}

LLR results related to the Moon's interior are of current interest.  These include the Love numbers, tidal dissipation $Q$ vs frequency, core dissipation, and core oblateness.  Core oblateness effects now appear to be significant.   In the future it may be possible to learn whether the Moon has a small solid core inside a liquid outer core.  There should be signatures in the rotation data if that is the case, but they could be small.  The 75 yr lunar pole wobble may be related to effects at the core/mantle interface \citep{Yoder_1981}.  High accuracy tracking might detect stimulation events for this free mode.  

Studies of gravitational physics look for perturbations of the orbit of the Moon and small relativistic effects in the time of flight to and from the Moon.  Currently, the most important contributions of LLR to gravitational physics are the equivalence principle test and the rate of change of the gravitational constant $G$.  In addition, LLR has sensitivity to geodetic precession and other effects of general relativity, and other future tests are expected. 
 
LLR contributes to geophysics and geodesy with Earth rotation, tidal acceleration, precession and nutation results.  Still, the existence of only two active stations is a limitation.  If LLR is to increase its impact in Earth sciences then more stations, with wide distribution, are needed.  

Looking toward the future, the analysis depends on high-quality data and improved range accuracy helps all results.  Lengthening data span strongly helps long-time-scale effects like station motion, Earth precession, 18.6 yr nutation, tidal acceleration, orbital inclination, node, and precessions, the search for changing $G$, and lunar core dissipation and free librations.  New retroreflectors on the Moon would most strongly help the lunar science, but would also benefit other areas.  The return to the Moon with robotic and manned spacecraft increases the importance of LLR contributions to science, ephemerides, and positions on the Moon while offering the opportunity for additional passive retroreflectors and active optical transponders.

%**************************************
\section*{Acknowledgments}
We acknowledge and thank the staffs of the Observatoire de la C\^ote d'Azur, Haleakala, and University of Texas McDonald ranging stations.  The research described in this paper was carried out at the Jet Propulsion Laboratory of the California Institute of Technology, under a contract with the National Aeronautics and Space Administration.

%***************************


\begin{thebibliography}{99}

% \bibitem[Names(Year)]{label} or \bibitem[Names(Year)Long names]{label}.
% (\harvarditem{Name}{Year}{label} is also supported.)
% Text of bibliographic item
%**************************************

\bibitem[Adelberger(2001)]{Adelberger_2001}
Adelberger, E. G., ``New Tests of Einstein's Equivalence Principle and Newton's inverse-square law,'' {\it Classical and Quantum Gravity, \bf 18}, 2397-2405, 2001.
 
\bibitem[Bertotti, Iess, Tortora(2003)]{Bertotti_Iess_Tortora_2003}
Bertotti, B., L. Iess, and P. Tortora, ``A test of general relativity using radio links with the Cassini spacecraft,'' {\it Nature, \bf 425}, 374-376, 2003.
 
\bibitem[Chapront, Chapront-Touz\'e, 
and Francou(2003)]{Chapront_etal_2003}
Chapront, J., M. Chapront-Touz\'e, and G. Francou, ``A new determination of lunar orbital parameters, precession constant and tidal acceleration from LLR measurements,'' {\it Astron. Astrophys., \bf 387}, 700-709, 2002.  

\bibitem[Damour and Nordtvedt(1993)]{Damour_Nordtvedt_1993}
Damour, T., and K. Nordtvedt, ``Tensor-scalar cosmological models and their relaxation toward general relativity,'' {\it Phys. Rev. D, \bf 48}, 3436-3450, 1993. 

\bibitem[Damour, Piazza, and Veneziano(2002)]{Damour_Piazza_Veneziano_2002}
Damour, T., F. Piazza, and G. Veneziano, ``Violations of the equivalence principle in a dilaton-runaway scenario,'' {\it Phys. Rev. D, \bf 66}, 046007, 1-15, 2002,  [arXiv:hep-th/0205111].

\bibitem[Davies, Colvin, and Meyer(1987)]{Davies_Colvin_Meyer_1987}
Davies, M. E., T. R. Colvin and D. L. Meyer, ``A unified lunar control network: The near side,'' {\it J. Geophys. Res., \bf 92}, 14177-14184, 1987.

\bibitem[Davies et al.(1994)]{Davies_etal_1994}
Davies, M. E., T. R. Colvin, D. L. Meyer, and S. Nelson, ``The unified lunar control network: 1994 version,'' {\it J. Geophys. Res., \bf 99}, 23211-23214, 1994.  

\bibitem[Davies and Colvin(2000)]{Davies_Colvin_2000}
Davies, M. E., and T. R. Colvin, ``Lunar coordinates in the regions of the Apollo landers,'' {\it J. Geophys. Res., \bf 105}, 20277-20280, 2000.  

\bibitem[Hood et al.(1999)]{Hood_etal_1999}
Hood, L. L., D. L. Mitchell, R. P. Lin, M. H. Acuna, and A. B. Binder, ``Initial measurements of the lunar induced magnetic dipole moment using Lunar Prospector magnetometer data,'' {\it Geophys. Res. Lett., \bf 26}, 2327-2330, 1999.  

\bibitem[Khan et al.(2004)]{Khan_etal_2004}
Khan, A., K. Mosegaard, J. G. Williams, and P. Lognonne, ``Does the Moon possess a molten core? Probing the deep lunar interior using results from LLR and Lunar Prospector,'' {\it J. Geophys. Res., \bf 109} (E9), 1-25, 2004.  

\bibitem[Konopliv et al.(1998)]{Konopliv_etal_1998} 
Konopliv, A. S., A. B. Binder, L. L. Hood, A. B. Kucinskas, W. L. Sjogren, and J. G. Williams, ``Improved gravity field of the Moon from Lunar Prospector,'' {\it Science, \bf 281}, 1476-1480, 1998.  

\bibitem[Konopliv et al.(2001)]{Konopliv_etal_2001} 
Konopliv, A. S., S. W. Asmar, E. Carranza, W. L. Sjogren, and D. N. Yuan, ``Recent gravity models as a result of the lunar prospector mission,'' {\it Icarus, \bf 150}, 1-18, 2001.  

\bibitem[Kuskov and Kronrod(1998)]{Kuskov_Kronrod_1998}
Kuskov, O. L., and V. A. Kronrod, Constitution of the Moon 5. ``Constraints on composition, density, temperature, and radius of a core, '' {\it Phys. Earth and Planetary Interiors, \bf 107}, 285-386, 1998.  

\bibitem[M\"uller, Nordtvedt, and Vokrouhlicky(1996)]{Mueller_Nordtvedt_Vokrouhlicky_1996}
M\"uller, J., K. Nordtvedt, Jr., and D. Vokrouhlicky, ``Improved constraint on the  1 PPN parameter from lunar motion,'' {\it Phys. Rev. D, \bf 54}, R5927-R5930, 1996.  

\bibitem[Nordtvedt(1968)]{Nordtvedt_1968}
Nordtvedt, K., Jr., ``Testing relativity with laser ranging to the Moon,'' {\it Phys. Rev., \bf 170}, 1186-1187, 1968. 
 
\bibitem[Nordtvedt(1970)]{Nordtvedt_1970}
Nordtvedt, K., Jr., ``Solar system E\"otvos Experiments,'' {\it Icarus, \bf 12}, 91-100, 1970. 
 
\bibitem[Nordtvedt, M\"uller, and Soffel(1995)]{Nordtvedt_Mueller_Soffel_1995}
Nordtvedt, K. L., J. M\"uller, and M. Soffel, Cosmic Acceleration of the Earth and Moon by Dark-Matter,'' {\it Astron. Astrophys., \bf 293}, L73-L74, 1995.  

\bibitem[Standish and Williams(2003)]{Standish_Williams_2003}
Standish, E. M., and J. G. Williams, Orbital Ephemerides of the Sun, Moon, and Planets,'' Chapter 8 of the {\it Explanatory Supplement to the American Ephemeris and Nautical Almanac}, in press, 2003.  

\bibitem[Williams, Newhall, and Dickey(1996)]{Williams_etal_1996}
Williams, J. G., X X Newhall, and J. O. Dickey, ``Relativity parameters determined from lunar laser ranging,'' {\it Phys. Rev. D, \bf 53}, 6730-6739, 1996.

\bibitem[Williams et al.(2001)]{Williams_etal_2001}
Williams, J. G., D. H. Boggs, C. F. Yoder, J. T. Ratcliff, and J. O. Dickey, ``Lunar rotational dissipation in solid body and molten core,'' {\it J. Geophys. Res., \bf 106}, 27933-27968, 2001. 

\bibitem[Williams and Dickey(2003)]{Williams_Dickey_2003}
Williams, J. G., and J. O. Dickey, ``Lunar Geophysics, Geodesy, and Dynamics,'' 13th International Workshop on Laser Ranging, October 7-11, 2002, Washington, D. C., eds. R. Noomen, S. Klosko, C. Noll, and M. Pearlman, NASA/CP-2003-212248, pp. 75-86, 2003.  {\tt http://cddisa.gsfc.nasa.gov/lw13/lw\underline{ }proceedings.html}  

\bibitem[Williams, Turyshev, and Murphy(2004)]{Williams_Turyshev_Murphy_2004}
Williams, J. G., S. G. Turyshev, and T. W. Murphy, Jr., ``Improving LLR Tests of Gravitational Theory,'' {\it International Journal of Modern Physics D \bf13}(3), 567-582, 2004, [arXiv:gr-qc/0311021]. ~ {\tt http://www.worldscinet.com/ijmpd/13/1303/S0218271804004682.html}

\bibitem[Williams, Boggs, and Ratcliff(2004)]{Williams_Boggs_Ratcliff_2004}
Williams, J. G., D. H. Boggs, and J. T. Ratcliff, ``Lunar Core and Tides.'' Abstract \#1398 of the Lunar and Planetary Science Conference XXXV, March 15-19, 2004. 
 
\bibitem[Williams, Turyshev, and Boggs(2004)]{Williams_Turyshev_Boggs_2004} 
Williams, J. G., S. G. Turyshev, and D. H. Boggs, ``Progress in lunar laser ranging tests of relativistic gravity,'' submitted to Phys. Rev. Lett., 2004.
  
\bibitem[Yoder(1981)]{Yoder_1981} Yoder, C. F., ``The free librations of a dissipative,'' {\it Moon, Philos. Trans. R. Soc. London Ser. \bf A}, 303, 327-338, 1981. 
 
\bibitem[Yoder(1985)]{Yoder_1995} Yoder, C. F., ``Venus' free obliquity,'' {\it Icarus, \bf 117}, 250-286, 1995.  

\end{thebibliography}
\end{document}